\documentclass[rsi,amsmath,amssymb,reprint]{revtex4-1}

\usepackage{graphicx}
\usepackage{dcolumn}
\usepackage{bm}

\usepackage[utf8]{inputenc}
\usepackage[T1]{fontenc}
\usepackage{mathptmx}

\begin{document}

\preprint{RSI/Shaposhnikov_2}

\title[A Powerful Pulsed “Point-Like” Neutron Source Based on the High-Current ECR Ion Source]{A Powerful Pulsed “Point-Like” Neutron Source Based on the High-Current ECR Ion Source}

\author{V.A. Skalyga}
 \altaffiliation[Also at ]{Lobachevsky State University \\ of Nizhny Novgorod, 603155 Nizhny Novgorod, Russia.}
\author{S.V. Golubev}
\author{I.V. Izotov}
\author{R.A. Shaposhnikov}
 \email[]{shaposhnikov-roma@mail.ru}
\author{S.V. Razin}
\author{A.V. Sidorov}
\altaffiliation[Also at ]{Lobachevsky State University \\ of Nizhny Novgorod, 603155 Nizhny Novgorod, Russia.}
\author{A.F. Bokhanov}
\author{M.Yu.~Kazakov}
\author{R.L. Lapin}
\author{S.S. Vybin}
\affiliation{ 
Institute of Applied Physics, Russian Academy of Sciences, 603950 Nizhny Novgorod, Russia}

\date{\today}

\begin{abstract}
The paper presents recent results of a “point-like” neutron source development based on a D-D fusion in a D-loaded target caused by its bombardment with a sharply focused deuterium ion beam. These developments are undergoing at the Institute of Applied Physics of Russian Academy of Sciences in order to study a possibility to create an effective and compact device for fast-neutron radiography. Last experiments with a beam produced by a gasdynamic high-current ECR ion source and its focusing with magnetic lens demonstrated that 60~mA of deuterium ions may be constricted to a transversal size of $\sim1$~mm at the focal plane. With a purpose to improve this result in terms of the beam current and its size a combined electrostatic and magnetic focusing system is proposed and analysed. It is shown that combined system may enhance the total beam current and reduce its footprint down to 0.13~mm. All numerical analysis was performed using the IBSimu code. 
\end{abstract}

\maketitle 

\section{Introduction}

A neutron tomography is one of perspective technologies in modern physics, which may be applied in medicine, archeology, engineering etc. The most commonly used neutronography with thermal neutrons requires specialized neutron sources, such as nuclear reactors and large accelerators equipped with long collimators, delivering thermal neutron fluxes with a small angular spread. However, these facilities are very expensive and complicated in operation, which hinders the spread of neutron tomography as a routine technique. Another approach for neutron imaging is a fast-neutron radiography \cite{anderson,adams_2016,adams_2015,andersson,dangendorf}. In this case there is no need for neutron thermalization and radiation from a point-like sources could be applied instead of parallel beams, which means higher efficiency of neutron utilization for an image production. For example, the evolution of X-ray tomography methods in general was connected with transition to the point-like radiation sources to increase efficiency and reduce minimal necessary dose \cite{anderson}. In frames of such technique the quality of the radiographic image increases with the decrease of the size of the radiation emitting region \cite{anderson}. In the present paper, some prospects of high-yield D-D neutron generator application for development of a point-like neutron source for fast-neutron radiography are discussed. In order to reach a high quality of the imaging system investigations on improvement of the ion beam focusing efficiency at the neutron producing target in case of high current density were performed.

The approach \cite{golubev_2017} used in the present research is based on the recent developments \cite{skalyga_2016_1,skalyga_2015,skalyga_2014} of the unique high-current ion source utilizing an ECR discharge sustained by powerful gyrotron radiation of millimeter wave range in open magnetic traps. The use of modern powerful high-frequency gyrotrons allows to increase the plasma density by at least one order of magnitude if compared to the conventional ECR ion sources (up to $10^{13} - 10^{14}$ cm$^{-3}$). A change in the plasma confinement regime happens with the plasma density increase – a so-called quasi-gasdynamic confinement \cite{skalyga_2016_1} is realized instead of a conventional collision-less regime. A distinctive feature of the quasi-gasdynamic confinement is a filled loss-cone in an electron velocity space. The plasma lifetime $\tau_e$ is then determined by a time needed for ions to escape the trap, which happens with the ion-sound velocity $V_s=\sqrt{T_e/M}$ ($T_e$ –- electron temperature, $M$ –- ion mass): $\tau_e=LR/V_s$ , where R, L –- magnetic mirror ratio and the trap length respectively. A typical value of the electron temperature is on the level of 20 -- 50~eV \cite{skalyga_2016_1}. A compact simple mirror magnetic trap with L=20~cm is used for plasma confinement, which yields $\tau_e\sim5$~$\mu$s lifetime (whereas the plasma lifetime is on the level of several ms in conventional ECR ion sources). A short lifetime together with high plasma density provides a possibility of ion beam formation with current density $j>10$~A/cm$^2$ (as $j\sim N_e/\tau_e$, $N_e$ being the plasma density). High-current ion beams with high quality, i.e. low emittance, extracted from the plasma of the ECR discharge confined in a simple mirror trap had been shown earlier \cite{skalyga_2015,skalyga_2014}. In particular, up to 500~mA of proton and deuteron beams with the normalized RMS emittance lower than 0.1~$\pi \cdot mm \cdot mrad$ were demonstrated. Furthermore, experiments in \cite{golubev_2019} showed the possibility of a point-like neutron source creation with the use of such ion beams, yielding the neutron flux of 10$^{10}$ s$^{-1}$.

This paper continues experiments described in \cite{golubev_2019} and presents results of investigation on a sharp focusing of the ion beam with magnetic lens, its comparison this numerical simulations and development of a more effective focusing systems. A combined focusing system consisting of both magnetic and electrostatic lenses is proposed. 

\section{Experiments}

\begin{figure*}
\includegraphics[width=1\linewidth]{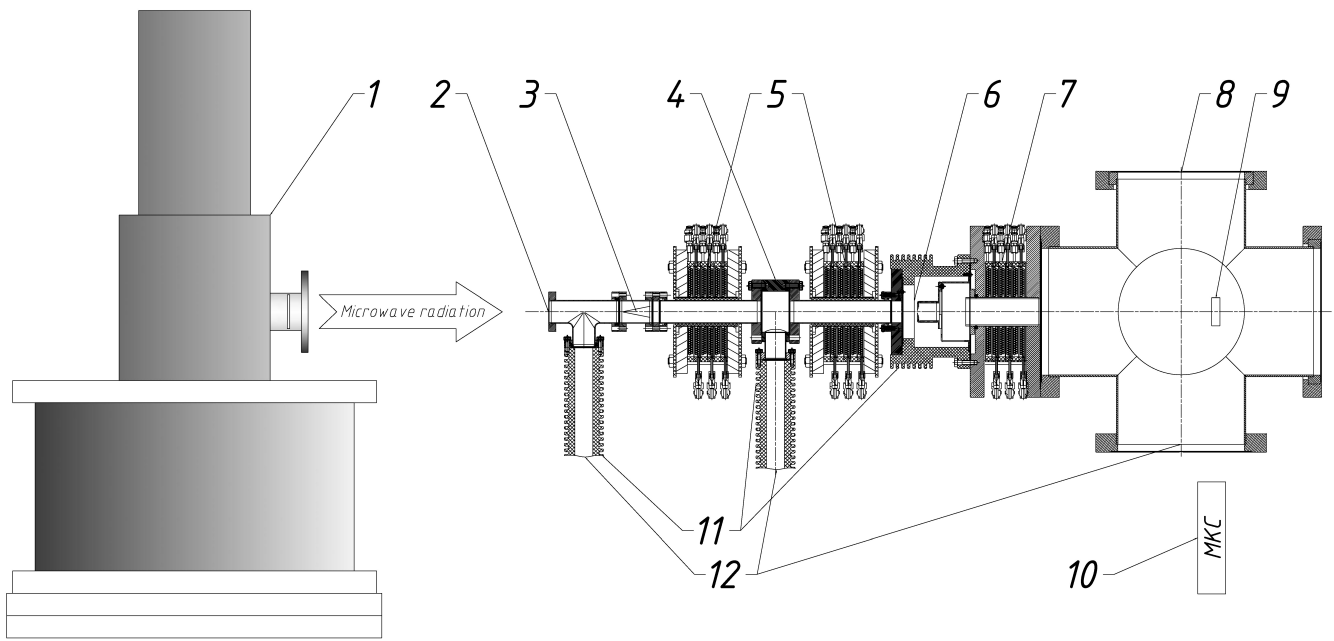}
\caption{\label{fig:scheme} 1 -- gyrotron, 2 -- quartz window, 3 -- microwave-to-plasma coupling system with embedded gas feed line, 4 -- plasma chamber, 5 -- pulsed magnetic coils, 6 - extraction system, 7 -– focusing magnetic lens, 8 -- diagnostic chamber, 9 -– Faraday cup, CsI scintillator or deuterium-loaded target, 10 -- neutron detector, 11 -- high voltage  insulators, 12 -- vacuum pumps.}
\end{figure*}

The experiments were performed at SMIS 37 (Simple Mirror Ion Source 37 GHz) experimental facility \cite{skalyga_2016_1,skalyga_2015,skalyga_2014}. The scheme of the experiment is shown in Fig.\ref{fig:scheme}.

A gyrotron (1) radiation with power up to 100 kW and frequency 37.5~GHz was used for a discharge start-up and plasma heating. The radiation was injected into a discharge vacuum chamber (4) quasi-optically through a quartz microwave window (2). An electromagnetic system (3) was used for the coupling of the waveguide mode to the plasma and to protect the quartz window from the plasma flux. A gas feeding system equipped with a pulsed gas valve was embedded into the microwave coupling system to provide neutral gas injection along the discharge axis. Magnetic field of the simple mirror configuration was produced by means of two pulsed magnetic coils (5). A two-electrode ion beam extraction system (6) with 5~mm aperture was placed 10~cm downstream the magnetic mirror. A newly designed high-voltage insulators (11) determined the maximum extraction voltage at the level of 100~kV. Vacuum pumps (12) provided residual gas pressure (measured between gas pulses) of $10^{-6}-10^{-7}$~Torr. A high-field magnetic lens (7) was used for the sharp focusing of the ion beam. Either a Faraday cup for the beam current measurements, a CsI scintillator for the ion beam shape visualisation, or the deuterium-loaded target for neutron production were placed downstream the magnetic lens at position (9). The scintillator luminescence was recoded with a MCP-CCD camera (not shown in the figure) through the axial optical flange of the diagnostic chamber (8), whereas the neutron flux measurements were performed with neutron detector (10) placed outside of the vacuum volume.

\begin{figure*}
\includegraphics[width=1\linewidth]{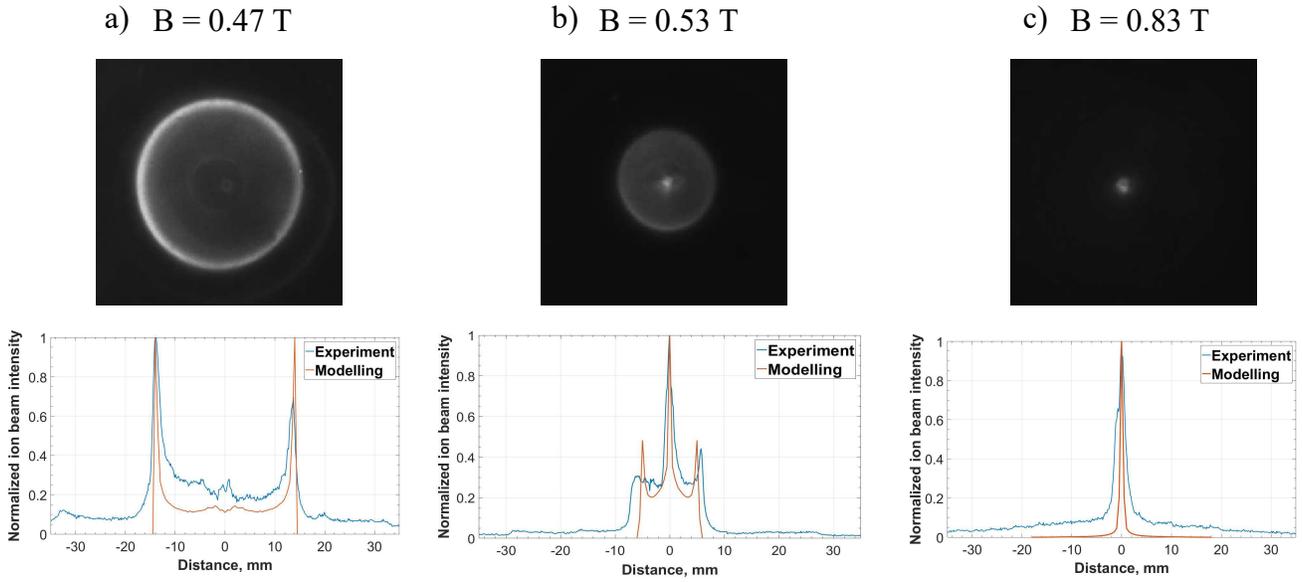}
\caption{\label{fig:CsI} CsI screen luminescence at different values of magnetic field in the lens centre (a) B = 0.47~T, (b) B = 0.53~T, (c) B = 0.83~T and comparison of experimental (blue) and modelled (orange) transverse beam intensity distribution.}
\end{figure*}

Before the experiments, the ion beam focusing at different fields of the magnetic lens was modelled using the open-source software IBSimu. Simulation results (orange curve) and experimentally obtained transverse beam profiles (blue curve) are demonstrated in Fig.\ref{fig:CsI}, where the CsI screen luminescence is also shown for a few values of magnetic lens field (measured at its center). The photographs and corresponding beam intensity distributions show that the transverse beam size decreases with the increase of the magnetic field from 28~mm at 0.47~T (Fig.\ref{fig:CsI}a) to $\sim$1 mm at 0.83~T (Fig.\ref{fig:CsI}c), being the smallest achieved one. Further increase of the magnetic field leads to an over-focusing of the beam at the scintillator position, thus enlarging its transverse footprint.

\begin{figure}
\includegraphics[width=1\linewidth]{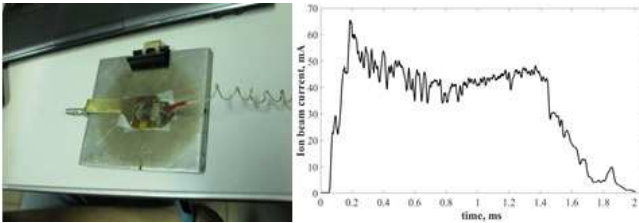}
\caption{\label{fig:target} TiD$_2$ target and ion beam current waveform.}
\end{figure}

The beam current value of 60~mA was measured with a movable Faraday cup under conditions of the ion source settings corresponding to the best focusing.  Keeping all the source settings fixed the TiD$_2$ target was placed at the ion beam focus for further measurements. A conformity of the settings was controlled by the beam current value collected by the target. The photograph of the target and a waveform of the ion beam current are shown in Fig.\ref{fig:target}.

\begin{figure}
\includegraphics[width=1\linewidth]{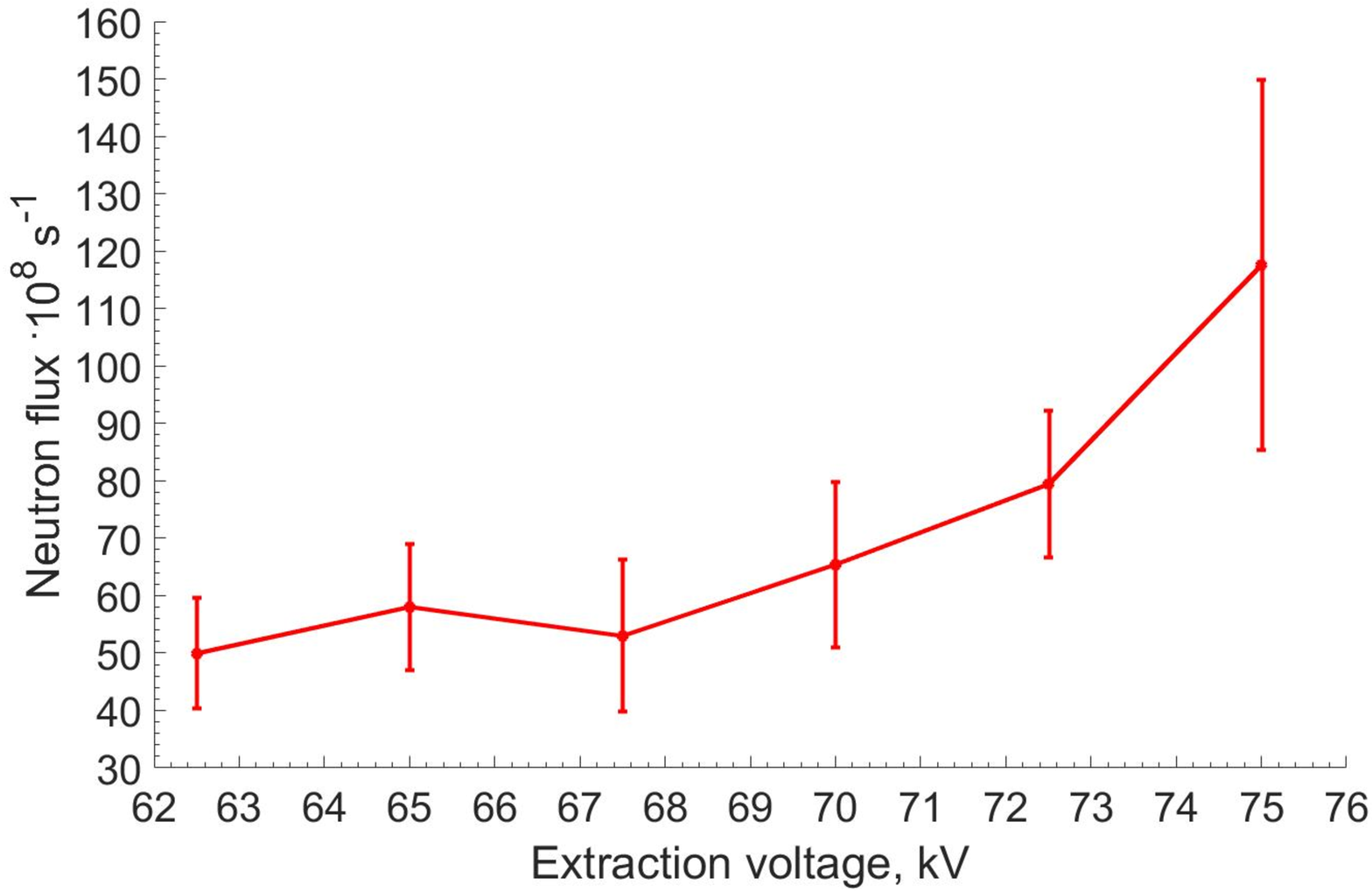}
\caption{\label{fig:yield} The dependence of neutron yield on the extraction voltage.}
\end{figure}

The experimentally obtained dependence of neutron yield on the extraction voltage using the dosimeter-radiometer placed outside the diagnostic chamber is shown in the Fig.\ref{fig:yield}. 

From the presented data it follows that the neutron flux increases with the extraction voltage growth according to the D-D reaction cross-section. The maximum ion beam acceleration voltage available at the experiment was limited to the value of 80~kV due to technical constraints. The detected neutron yield reached the value of 10$^{10}$ s$^{-1}$ at the beam energy level of 75 -- 80~keV. Further increase of the neutron yield may be achieved by using of a wider plasma electrode aperture, i.e. 1~cm, which would lead to an increase in the ion beam current up to the value of a few hundred of mA. However, the increase in the extraction aperture could lead to significant losses of the beam, as its transversal size would not fit the lens aperture due to the beam divergence. To increase the current delivered to the target it is suggested to reduce beam divergence by means of an additional focusing electrode in the extraction system prior to the magnetic lens. The combined focusing system was studied in frames of numerical simulations, corresponding results are shown below.

\section{Modelling of the combined ion beam focusing system}

\begin{figure}
\includegraphics[width=1\linewidth]{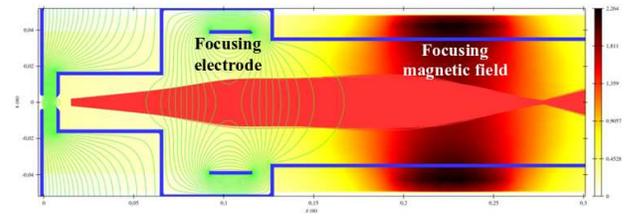}
\caption{\label{fig:transport} Ion trajectories (red lines), equipotential surfaces (green lines) and the magnetic field distribution (gradient color, T) in the entire beam transport region.}
\end{figure}

As it has been mentioned above, it is suggested to combine electrostatic and magnetic lens to reduce the ion beam transverse size and increase its current at the target. Additional electrostatic focusing may be placed close to the beam formation region, thus allowing to reduce the beam divergence where its transverse size is yet small, thus forming paraxial beam. The results of numerical modeling are presented in the Fig.\ref{fig:transport}. The parameters used in simulation are the following: extraction voltage is 60~kV (the focusing electrode being at the same potential), magnetic field in the magnetic lens center is 1.8~T, beam space charge compensation is 95\%, initial current density at the plasma electrode aperture is 650~mA/cm$^2$, electron temperature is 10~eV, ion temperature is 1~eV.

\begin{figure}
\includegraphics[width=1\linewidth]{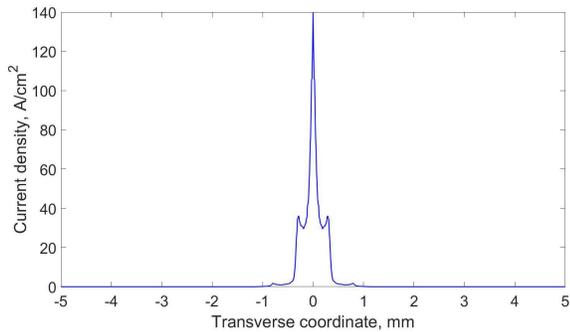}
\caption{\label{fig:distr} The ion beam current density distribution.}
\end{figure}

The ion beam current density transversal distribution at the focal plane corresponding to the data plotted in Fig.\ref{fig:transport} is shown in the Fig.\ref{fig:distr}. The ion beam full width at half maximum is as low as 0.13~mm.

\begin{figure}
\includegraphics[width=1\linewidth]{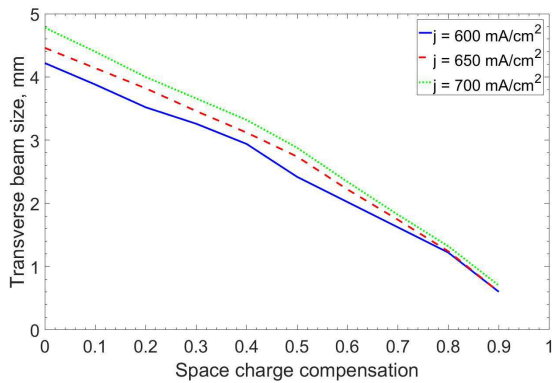}
\caption{\label{fig:beamsize} The dependence of transverse beam size on the space charge compensation at different values of current density: j=600~mA/cm$^2$ (blue solid curve), j=650~mA/cm$^2$ (red dashed curve),  j=700~mA/cm$^2$ (green dotted curve). Beam size was measured at the level of 10\% from the maximum.}
\end{figure}

As far as presented calculations were done for 95\% of the beam space charge compensation (SCC), additional studies of this parameter influence on focusing efficiency were performed.  The results of numerical simulations of ion beam transverse size dependence on the SCC at different values of current density are presented in Fig.\ref{fig:beamsize}.

The dependence demonstrates that the SCC growth leads to the decrease of the ion beam size. It could be explained by the fact that the addition of electrons to the ion beam reduces the effect of Coulomb repulsion forces. Thus, the ion beam focusing could be improved significantly by providing high SCC. The Fig.\ref{fig:beamsize} shows a decrease of the beam spot size by seven times when the SCC increases from zero to 0.9. Instead of full width at half maximum, the data presented in Fig.\ref{fig:beamsize} uses the term ``beam size'', which is the beam width calculated at the level of 10\% from the maximum intensity to make the analysis easier in case of a ring shape of the beam. It must be mentioned that the data in Fig.\ref{fig:beamsize} are calculated at a certain longitudinal position selected as a point of the best focusing in case of the maximum SCC. With the change of SCC the shift of the best focusing point along the axis is observed. This is mainly because in the magnetic lens the ion trajectory convergence depends on its transversal coordinate, and the SCC level strongly affects which part of the beam contributes the most to the current density at the focal spot. This effect is rather complex, though the data in Fig.\ref{fig:beamsize} gives a good estimate for the central part of the beam being focused. It is also seen from the dependences that beams with high current density are harder to focus at the low SSC, but with the increase of the SSC the difference in the beam size decreases.

\section{Conclusion}

The results reported here demonstrate the prospects of the suggested approach. The smallest achieved transverse size of the ion beam was $\sim$1 mm at the focusing magnetic field of 0.8~T. The total deuterium beam current delivered to the target reached 60~mA at 80~keV of its energy, yielding the neutron flux of 10$^{10}$~s$^{-1}$. The comparison of numerical simulations with experimental data have shown their good agreement and proven the modelling reliability for further developments. It was demonstrated that the use of combined electrostatic and magnetic focusing system may significantly decrease the ion beam diameter (down to 0.13~mm) and increase the ion current delivered to the target as a result of reduced beam losses because of its divergence reduction. So, adding the focusing electrode may noticeably enhance the experimental results. 

Obtained values of the neutron flux are already applicable for experimental tests on neutron imaging. Further research will be aimed at the studies of the combined focusing of a deuterium beam and obtaining the first neutronographic images of test objects.

It should be mentioned that the presented neutron yield values corresponds to its intensity during the microwave pulse. As far as our experiments were conducted in the pulsed mode with 0.1~GHz repetition rate the time-averaged neutron flux was lower by 4~orders of magnitude. To avoid any misunderstanding we would like to stress the reader’s attention that such low repetition rate is not a kind of any principal limitation for this type of neutron generator, rather being a restriction caused by the gyrotron power supply. Its improvement together with installing of a permanent magnet plasma trap (both are commercially available) would allow us to raise the repetition rate up to 100~Hz or even switch to a continuous wave operation. In this case it should be noted that the heat load to the target will become much higher and would be the most problematic point. This problem could be solved by using of a rotating target. Thus, we conclude that with a few facility technical improvements the time- averaged neutron flux could be increased up to the values required in the neutron radiography methods.

To compare the presented results with performance of existing systems for fast-neutron tomography reported by other authors it is reasonable to refer to pure scientific studies \cite{adams_2016,adams_2015,andersson} and commercial developments \cite{phoenix,adelphi}. In \cite{adams_2016,adams_2015,andersson}neutron generators with a total yield of $10^7-10^8$ neutrons per second were used and that is much lower than the suggested generator based on high-current gasdynamic ion source could provide. Improvement of experimental schemes used in \cite{adams_2016,adams_2015,andersson} with sufficiently more powerful neutron source could lead to a very important results for fast-neutron tomography evolution. An absolutely different situation could be observed around commercial devises. In \cite{phoenix,adelphi} neutron generators with a total yield up to 10$^{11}$ are announced, moreover Phoenix Nuclear Labs claims that their Alectryon neutron generator is the most powerful one in the world. However, authors were not able to find the published technical description to estimate a possible neutron flux density at the irradiated object, also the emitting region size (i.e. imaging quality) is unknown. Thus, only indirect data could be used for the comparison. A value of maximum available ion beam current could be used in such a way. In \cite{phoenix} it is shown that the maximum current of the Alectryon ion source is 90~mA. In the present paper we have discussed the approach of the combined focusing system which could allow to use the whole potential of the high current ECR ion sources able to produce high quality beams with 500~mA current. This means that the proposed system could be much more powerful than the competitive solutions after some improvements.

\begin{acknowledgments}
The work was supported by the project of the Russian Science Foundation \# 16-19-10501.
\end{acknowledgments}

\bibliography{ms}

\end{document}